\documentclass{article}
\usepackage{spconf,amsmath,graphicx}
\usepackage{booktabs}
\usepackage{subfigure}
\usepackage{multirow}
\usepackage{url}
\usepackage{bm}
\usepackage{algorithm}
\usepackage{algorithmic}
\usepackage{CJKutf8}

\title{Dilated Convolutional Neural Network-Based Deep Reference Picture Generation for Video Compression}
\name{Haoyue Tian$^{1}$, Pan Gao$^{1}$, Ran Wei$^{2}$, Manoranjan Paul$^{3}$ \thanks{This work is supported by Aeronautical Science Foundation of China under Grant 201951052001.} }
\address{
	$^{1}$ Nanjing University of Aeronautics and Astronautics, China \\ 
	$^{1}$ \{tianhy, pan.gao\}@nuaa.edu.cn \\
	$^{2}$ Science and Technology on Electro-optic Control Laboratory, China \\
	$^{3}$ School of Computing and Mathematics, Charles Sturt University, Australia
	\vspace{-2mm}
	}

\begin{document}
\ninept
\maketitle

\begin{abstract}
Motion estimation and motion compensation are indispensable parts of inter prediction in video coding. Since the motion vector of objects is mostly in fractional pixel units, original reference pictures may not accurately provide a suitable reference for motion compensation. In this paper, we propose a deep reference picture generator which can create a picture that is more relevant to the current encoding frame, thereby further reducing temporal redundancy and improving video compression efficiency. Inspired by the recent progress of Convolutional Neural Network(CNN), this paper proposes to use a dilated CNN to build the generator. Moreover, we insert the generated deep picture into Versatile Video Coding(VVC) as a reference picture and perform a comprehensive set of experiments to evaluate the effectiveness of our network on the latest VVC Test Model--VTM. The experimental results demonstrate that our proposed method achieves on average 9.7\% bit saving compared with VVC under low-delay P configuration. 
\end{abstract}
%

\section{Introduction}
\label{sec:intro}
In classical block-based hybrid video coding framework, inter prediction is the core technology to eliminate temporal redundancy. The basic idea of inter prediction is to use a matching block in the reference picture to predict the current block. During prediction, the motion vector (MV) and residual error between them are encoded. Since the movement of the object has a certain continuity, the motion of the same target between two frames may not be carried out in the unit of integer pixel. That is, the matching block may be located at fractional pixel positions in the reference picture. However, fractional pixels do not exist, which thus need to be interpolated by integer pixels. Usually, it uses row- or column-adjacent integer pixels through the linear calculation to get fractional pixels. 

\begin{figure}[ht]
	
	\centerline{\includegraphics[width=0.45\textwidth]{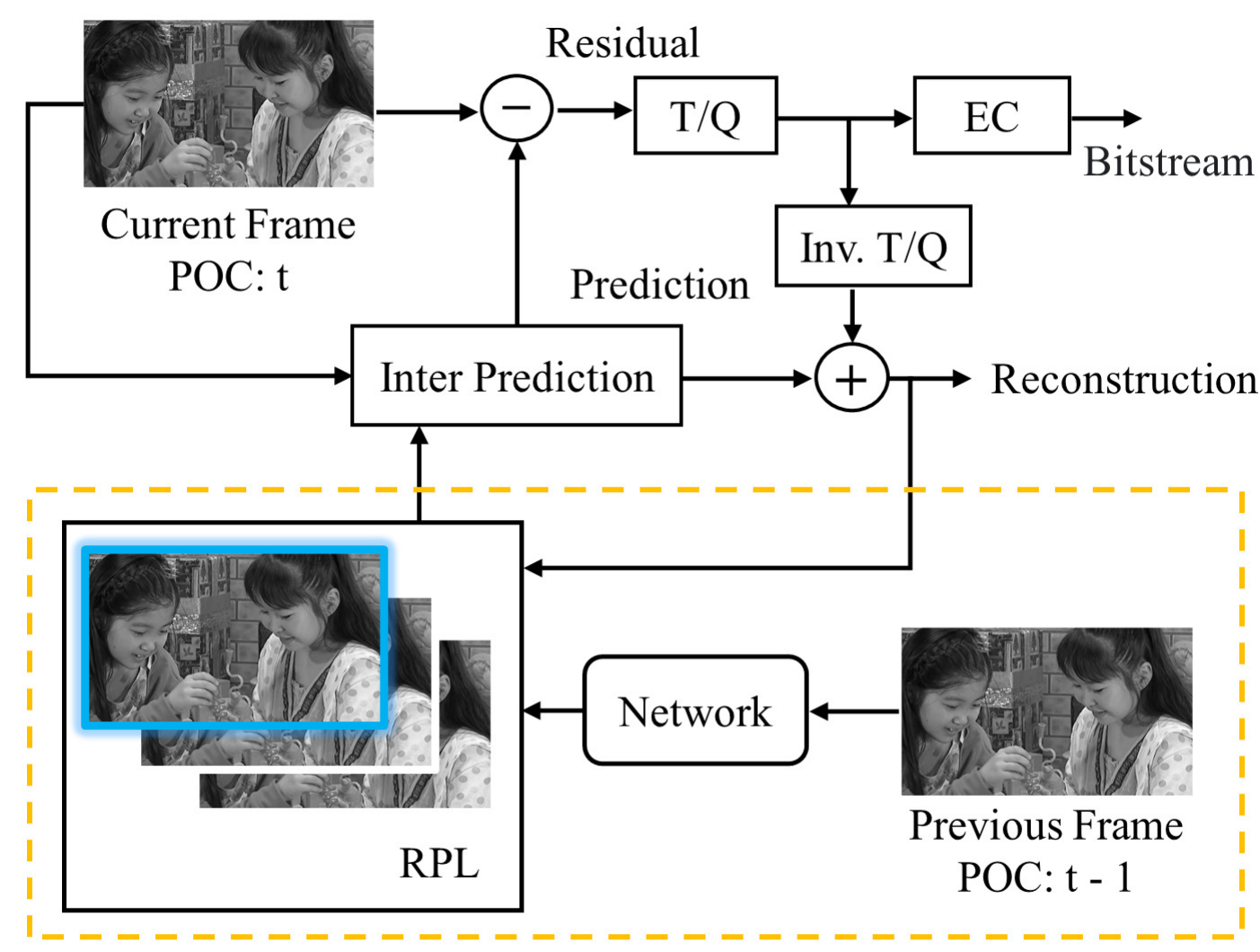}}
	\vspace{-3mm}
	
	\caption{Overview of the encoding process. T represents transformation, Q represents quantization, entropy coding is marked as EC, and RPL is the reference picture list. In RPL, the picture with a blue border is the network output, which is used to replace the first original reference picture in the buffer.}
	\label{overview}
	\vspace{-3mm}
\end{figure}

In H.264/AVC \cite{H264}, the prediction values at half-sample positions are obtained by using a one-dimensional 6-tap filter horizontally and vertically. In High Efficiency Video Coding (HEVC) \cite{HEVC} and Versatile Video Coding (VVC) \cite{VVC}, a symmetric 8 tap filter for half-sample interpolation and an asymmetric 7 tap filter for quarter-sample interpolation are included \cite{Interpolation}. However, considering the richness of natural videos, the traditional fixed fractional pixel interpolation filter may not be able to handle different types of contents well enough.

\begin{figure*}[t]
	\centerline{\includegraphics[width=0.8\textwidth]{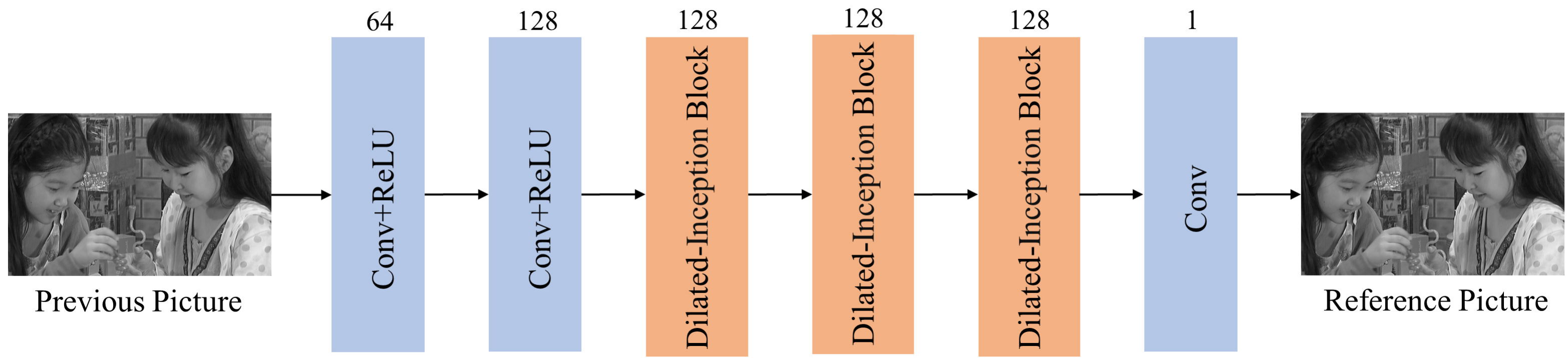}}
	\vspace{-3mm}
	\caption{Schematic illustration of the proposed network, which consists of a series of convolutional modules and dilated-inception blocks. Parameters on each layer indicate the output channel number of current layer.}
	\label{network}
	\vspace{-3mm}
\end{figure*}

\begin{figure}[t]
	\centerline{\includegraphics[width=0.45\textwidth]{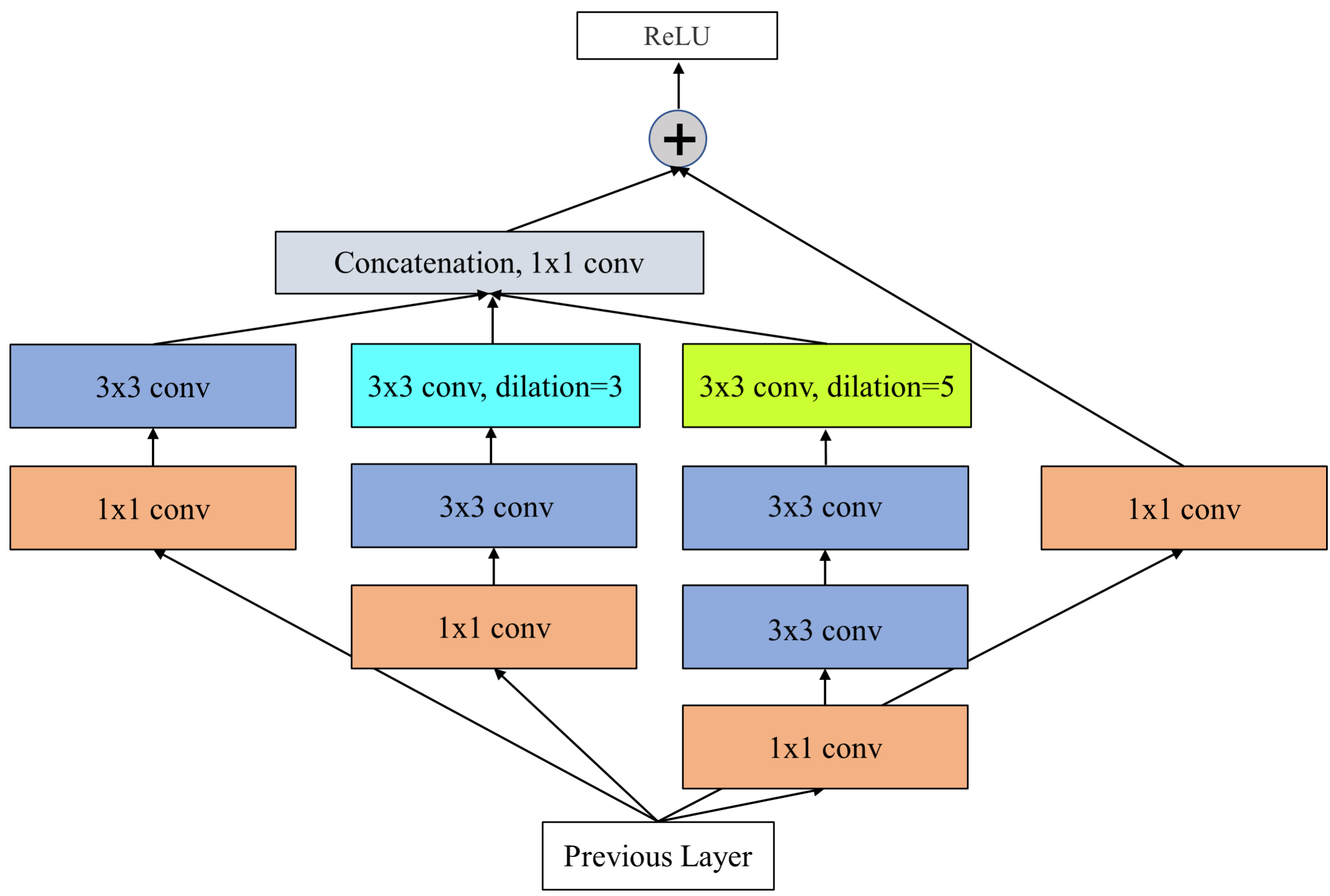}}
	\vspace{-2mm}
	\caption{Specific structure of the proposed Dilated-Inception block, consisting of three convolution branches and an approximate identity branch.}
	\label{block}
	\vspace{-4mm}
\end{figure}

Recently, convolutional neural network and deep learning have achieved great success in image processing tasks. Dong \emph{et al.} proposed a convolutional neural network, called Super-Resolution Convolutional Neural Network (SRCNN) \cite{SRCNN}, which directly learned an end-to-end mapping between low- and high-resolution images. Dai \emph{et al.} proposed the so-called Variable-Filter-Size Residue Learning Convolutional Neural Network (VRCNN) \cite{VRCNN} using multiple kernel size filters in one layer, which had the advantage to get more feature information with different receptive field sizes. Dilated Convolution was utilized in Yu \emph{et al.} \cite{Receptive} for object detection, which demonstrated that dilated convolution can enlarge the receptive field to capture multi-scale information. 

The excellent performance of CNN in image processing tasks provides some insights for video compression and coding \cite{DLBVC}, \cite{RDIP5}. 
Many previous works have showed the possibility of rate-distortion (RD) performance improvement by using deep learning in the sub-process during coding, e.g., block up-sampling \cite{RDIP2}, in-loop filtering \cite{RDIP3}, fractional interpolation \cite{RDIP4}, deep reference interpolation of B frame \cite{EMC}, etc. Liu \emph{et al.} proposed a deep frame interpolation network \cite{drgw} to generate additional deep reference pictures for B frames.

Inspired by those works, this paper proposes a dilated convolutional neural network to generate reference pictures for further improvement in inter prediction. We have made the following technical contributions in this paper:
\begin{enumerate}
	\item We propose a generator for P frame, which is more challenging compared to that for B frame as only one reference frame is available to utilize. To this end, we generate the deep reference for P frame using a dilated convolutional neural network. 
	With use of the proposed network model, we can extract more accurate temporal feature, and then provide a better reference.
	\item We propose to integrate the generated reference picture into the latest VVC codec. At the encoder of VVC, we substitute the original reference picture with the network output, which has less temporal redundancy due to end-to-end learning, thus improving the compression performance. 
\end{enumerate}

Fig.~\ref{overview} shows the video compression process. The upper part of the figure is the inter prediction process in VVC without integration of our method. The bottom of the figure is the process of inserting a reference picture generated by our method into the VVC, shown by a yellow dash-line box as depicted in Fig.~\ref{overview}. We can see that after each frame is encoded, its reconstructed picture is retained at the encoder, to serve as a reference picture for encoding the following frames. When encoding frame $t$ (which is not the I frame) under low-delay P (LDP) configuration, we use the previous frame, the frame $t-1$, as the input to the network, and then the network output, which is shown with the blue border in Fig.~\ref{overview}, is used to replace the original reference picture. Finally, the encoder selects the best reference picture from the list for inter prediction. 

The rest of this paper is organized as follows: Section \uppercase\expandafter{\romannumeral2} presents our network structure and some training details. Then we discuss how to integrate our reference picture into VVC in Section \uppercase\expandafter{\romannumeral3}. Section \uppercase\expandafter{\romannumeral4} presents some results and Section \uppercase\expandafter{\romannumeral5} concludes this paper.


\section{Network Training}
\vspace{-2mm}
\label{sec:format}

In this section, we will describe the specific structure of our proposed dilated convolutional neural network and the details of the training process of the network.

\vspace{-2mm}
\subsection{Network Structure}

We design the network structure according to CNN-based artifact removal \cite{VRCNN} and dilated convolution \cite{Receptive}. The overall scheme of the network is shown in Fig.~\ref{network}. 
Note that in training, we divide the picture into blocks, and each block can find a matching block in the previous picture. We thus optimize the model to map the matching block in the previous picture to its associated block in the current picture. And in testing, the previous picture is directly taken as input to the network, and the output is the reference picture to be used for predicting the current picture.

Input data goes through two convolution layers at first, and ReLU \cite{ReLU} is added to each convolution layer as an activation function. After that, three dilated-inception blocks are added. Finally, a $3\times3$ convolution layer is used in the last layer of the network to generate the final output. For each block, as shown in Fig.~\ref{block}, we use the inception module \cite{Inception} as the basic structure because we expect the block can obtain multi-scale feature map information from the previous layer. Then to acquire much contextual information, we apply dilated convolution.
We add it to the block and set its dilation rate to enlarge the size of the dilation, so that the range of the receptive field can be expanded without losing the resolution of the feature map.

For each branch, we add a $1\times1$ convolutional layer at first, the main purpose of which is to reduce the dimension to decrease convolution parameters while keeping spatial resolution unchanged. Then, we add standard convolution and dilated convolution to the first three branches. In this first branch, we use the standard $3\times3$ convolution. For the second branch, we use both standard convolution and a dilated convolution with a rate of 3. In the third branch, we use two standard $3\times3$ convolutions and one dilated convolution with a rate of 5. Note that, two stacked $3\times3$ convolutions are equivalent to a $5\times5$ convolution in terms of capturing receptive field but having fewer parameters. In this design, the receptive field sizes for the outputs in these three branches are 3, 9, and 15, respectively. Then, we concatenate the outputs of the three branches, aiming to combine the information obtained from different receptive fields. 
On the rightmost branch, we employ only a simple $1\times1$ convolutional layer. Consequently, the output obtained by this branch, to a large extent, still carries the information of the original input feature map.

Finally, we use the weighting operation to combine the feature maps after concatenation on the left and the feature maps on the right:

\begin{equation}
	X_{l+1} = ReLU \left[ k\cdot\left(F^1 * \varphi\right) + F^2 * X_{l} \right],
	\label{weight}
\end{equation}
where $\varphi$ is the feature maps after concatenating operation, $F^1 * \varphi$ and $F^2 * X_{l}$ are the outputs of the $1\times1$ convolutional operation on concatenated feature maps $\varphi$ and on the feature maps in the previous layer $X_{l}$ respectively, shown in Fig.~\ref{block}, and $k$ represents the proportionality factor in the range $[0,1]$, which determines how much features learned at this layer are preserved.

\begin{figure}[t]
	\centerline{\includegraphics[width=0.47\textwidth]{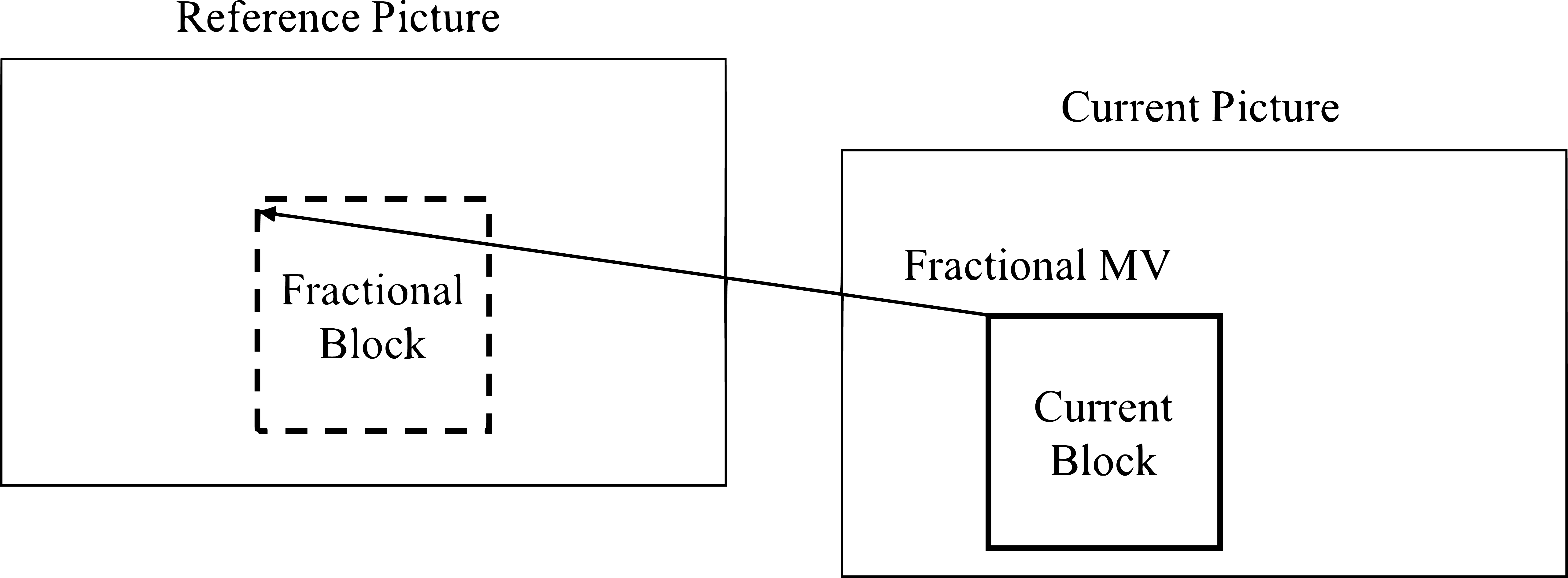}}
	\vspace{-3mm}
	\caption{Illustration of the process of generating training data for training the network.}
	\label{dataset}
	\vspace{-4mm}
\end{figure}

\vspace{-2mm}
\subsection{Training Details}

Since VVC is encoded based on block, we also divide the picture into small blocks during network training. Besides, choosing the blocks as the training samples rather than whole images can have the advantage of reducing the training complexity and augmenting training data. First, we select two consecutive reconstructed frames as the reference picture and the current picture, respectively, so the luminance change of the same target in these two frames is very small, and the movement of the same target between two consecutive frames is also small. 
It is assumed that the pixels of a block have the same motion trace, and hence we decide to use the Lucas-Kanade optical flow method \cite{LKOF} to derive fractional MV. 

Next, the training data are obtained through the process shown in Fig.~\ref{dataset}. In the current picture, current block is marked as ground truth (Y) of the network, then the position of the fractional block in the reference picture, depicted by the red dash line in Fig.~\ref{dataset}, is found through the fractional MV. Since the fractional pixel has no actual pixel value, it is necessary to find the corresponding position of the integer block by moving the referenced fractional block towards the top-left direction until the nearest integer pixels \cite{FPMC}, depicted by the blue line in Fig.~\ref{dataset}. This integer block is marked as input (X), so the pair of (X, Y) is used as a training sample for the network model. 

\newcommand{\tabincell}[2]{\begin{tabular}{@{}#1@{}}#2\end{tabular}} 

\begin{table}[t]
	\centering
	\caption{PSNR(dB) results of the networks trained with different block sizes under common test sequences.}
	\scalebox{0.83}{
		\begin{tabular}{cccccc}
			\toprule
			& \bm{$16\times16$} & \bm{$24\times24$} & \bm{$32\times32$} & \bm{$40\times40$} & \bm{$48\times48$} \\
			\midrule
			PartyScene & 31.6294 & 31.8747 & \textbf{32.3621} & 29.3443 & 26.7134 \\
			BQSquare & 27.4042 & 27.4983 & \textbf{27.828} & 26.0837 & 24.2715 \\
			FourPeople & 34.5303 & 34.6106 & \textbf{34.6108} & 34.259 & 33.6407 \\
			ChinaSpeed & 22.8142 & 22.769 & \textbf{22.8596} & 22.4953 & 21.9094 \\
			\bottomrule
		\end{tabular}
	}
	\label{tab1}
	\vspace{-4mm}
\end{table}

We consider the entire network as a mapping function $F$ and learn the network parameters $\theta$ through minimizing the loss $L(\theta)$ between the predicted blocks $F(X; \theta)$ and the corresponding ground truth $Y$. We use Mean Squared Error (MSE) as the loss function:

\begin{equation}
	L(\theta) = \frac{1}{M}\sum\limits_{i=1}^M \left[ \frac{1}{m\times n} \left \| F \left(X; \theta)-Y \right)\right \|_{F}^2 \right],
	\label{loss}
\end{equation}
where $M$ is the number of training samples, $m$ and $n$ represent the width and height of the block, respectively. We initially set the learning rate to $10^{-4}$ and adjust the learning rate at equal intervals. In addition, the optimizer for the network is Adadelta \cite{ADA} and the minibatch size is 32. After training for around 80 epochs, the training loss has converged.

\begin{table}[t]
	\vspace{-2mm}
	\centering
	\caption{PSNR(dB) of deep references generated by various models. }
	
	\scalebox{1.08}{
		\begin{tabular}{cccc}
			\toprule
			& VRCNN & VRF & Proposed  \\
			\midrule
			PartyScene & 24.1487 & 25.6505 & \textbf{26.0894} \\
			FourPeople & 29.8353 & 34.5041 & \textbf{34.673} \\
			BQSquare & 23.1371 & 26.922 & \textbf{27.6708} \\
			BasketballDrillText & 23.5848 & 24.8023 & \textbf{24.9879} \\

			\bottomrule
		\end{tabular}
	}
	\label{tabpsnr}
	\vspace{-4mm}
\end{table}

\section{Integration Into VVC}
During the encoding process of the VTM, mode decision is performed on the current Coding Unit (CU). VTM will check the various modes of intra-prediction and inter-prediction. 
For the inter prediction mode under LDP configuration, before encoding a frame, a forward reference picture list is constructed, which stores some reconstructed pictures of previous frames that have been encoded. Encoder iterates the search on these candidate pictures, finally selecting the block in the reference picture with the minimum distortion as the reference block for the coding block in the current frame.

\begin{table}[!t]
	\vspace{-2mm}
	\centering
	\caption{SSIM of deep references generated by various models. }
	
	\scalebox{1.1}{
		\begin{tabular}{cccc}
			\toprule
			& VRCNN & VRF & Proposed  \\
			\midrule
			PartyScene & 0.9449 & 0.955 & \textbf{0.957} \\
			FourPeople & 0.9922 & 0.997 & \textbf{0.9971} \\
			BQSquare & 0.9757 & 0.9891 & \textbf{0.9909} \\
			BasketballDrillText & 0.888 & 0.8962 & \textbf{0.8973} \\

			\bottomrule
		\end{tabular}
	}
	\label{tabssim}
	\vspace{-2mm}
\end{table}

Our approach is to use the previous frame of the current frame as the network input, and the purpose is to output a picture that is more similar to the current frame through the trained network. Then, we replace the picture in the original reference list with the model predicted one. We will discuss the compression results in the Section below.

\vspace{-2mm}
\section{Experimental Results}

\subsection{Experimental settings}

In this paper, we use the PyTorch framework to implement our model. For the training data, since we train the network on the basis of block as mentioned above, we use one common video sequence, namely \verb|BlowingBubbles|, to generate blocks. The strategy of using blocks instead of the whole image for training has been demonstrated very useful in high-resolution image restoration task \cite{FPMC}, \cite{3DVC}. 
Through experiments, it is found that the size of block affects the performance of the network.
In order to reasonably select the block size, we train the network with sizes of $16\times16$, $24\times24$, $32\times32$, $40\times40$ and $48\times48$ respectively, and test the trained network in different classes of HEVC test sequences. Some results are shown in Table~\ref{tab1}.
We found that the network output obtained by setting the block size to $32\times32$ is closer to ground truth than other sizes. 
Therefore, we choose this size to create our data set, which finally consists of more than 42000 block pairs.

For the encoder of VVC reference software VTM (version 10.0) \cite{VTM}, we have incorporated the proposed algorithms into it and compared our scheme with the unaltered VTM. We follow the VVC common test conditions and use LDP configuration to do compression performance test, under 4 quantization parameters (QPs): 22, 27, 32, and 37 \cite{QP}. In addition, since the human visual system is less sensitive to movement of chrominance than to luminance, video compression aims to store more luma details. Therefore, this paper focuses on the luminance component, and experiments on chroma are straightforward. 
Finally, BD-rate \cite{BDRate} is calculated to quantify the bit saving of different schemes relative to the original VTM.
\vspace{-2mm}

\subsection{Validation of the proposed model}
According to the previous analysis, the more similar the reference picture generated by the network to the current coding frame, the higher the coding efficiency will be. Therefore, in order to verify the effectiveness of our model, we compared the PSNR and SSIM of the deep reference pictures generated by three models (VRCNN \cite{VRCNN}, VRF \cite{EMC} and our model). Note that VRF model is a network that solves the problem of B frame interpolation, that is to say, the purpose of the network is to input two frames to generate an intermediate frame. To be fair, we modified their network structure to adapt it to a P frame generator that uses the previous picture to get the next one. 

\begin{figure}[t]
	\centerline{\includegraphics[width=0.48\textwidth]{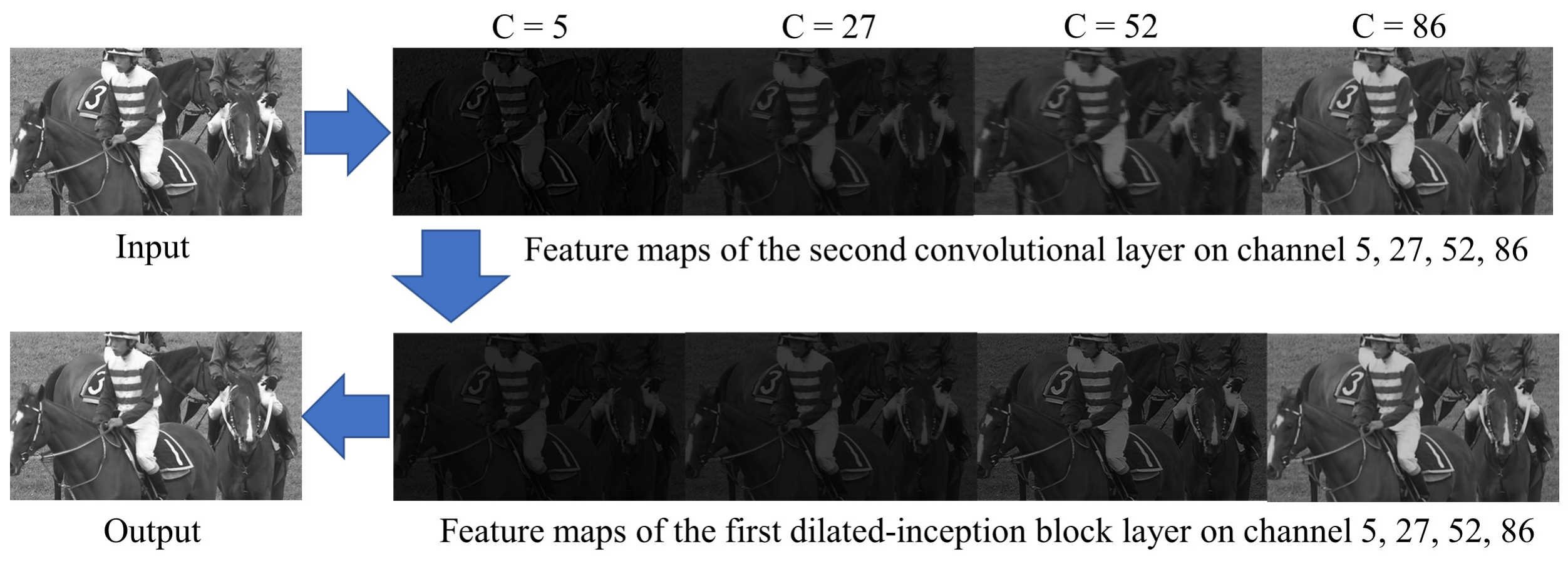}}
	\vspace{-2mm}
	\caption{Visualization of feature maps of different hidden layers.}
	\label{feature}
\end{figure}

We selected some sequences from the four classes of test sequences. In a GOP, experiments were performed on all frames except for the I frame. The average PSNR results and the average SSIM results are shown in Table~\ref{tabpsnr} and Table~\ref{tabssim}, respectively. Furthermore, we extracted some of the feature maps from some hidden layers to visualize how the network works, as shown in Fig.~\ref{feature}. We found that the feature maps after going through the dilated inception block can better show the details of the moving object.

\begin{figure}[t]
	\centerline{\includegraphics[width=0.47\textwidth]{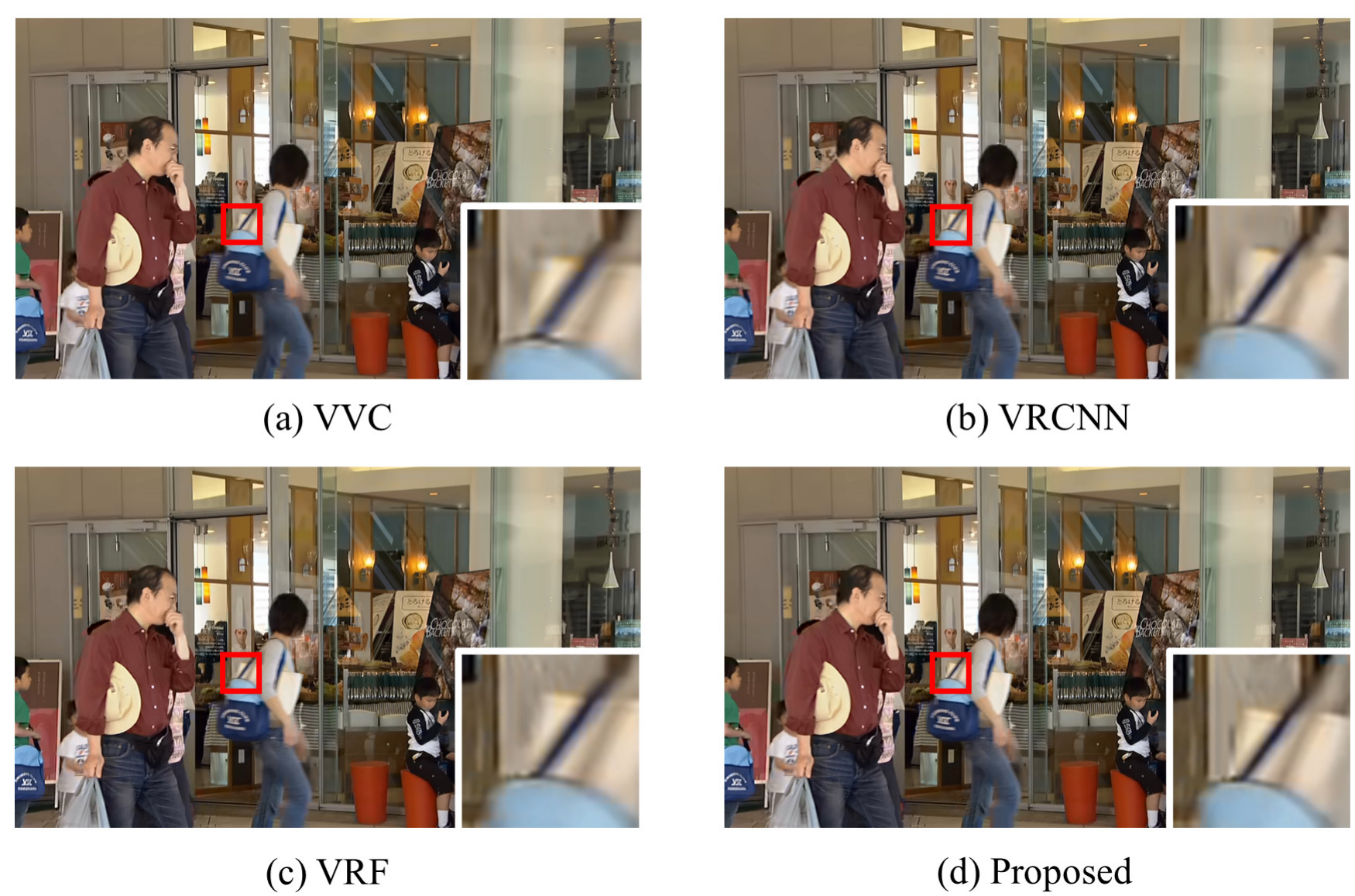}}
	\vspace{-2mm}
	\caption{The sixth frame from BQMall under QP: 32.}
	\label{example}
	\vspace{-4mm}
\end{figure}

\vspace{-2mm}
\subsection{Coding Results and Analysis}
We also compared our model with other models when used in video coding. For all network models, we use the same method to get reference pictures. Table~\ref{tab2} summarizes the BD-rate results of our scheme and other methods compared to the original VVC evaluated on the Y component. It can be observed that our proposed method leads to significant bit saving, which is on average 9.7\%. 
According to the results, we believe that the network we trained has a better impact on the reference picture prediction for natural video sequences compared with traditional VVC method and other two state-of-art methods. 
Note that the BD-rates for ChinaSpeed and SlideEditing sequences are positive, which means that the deep reference frames generated by the models are not as good as the original reference frame. This is probably because, these two sequences are screen content videos, whose characteristics (e.g., the content and motion intensity) are quite different from the natural video sequence used for training, and our trained model exhibits limited transferability.  

\begin{figure}[t]
	\vspace{-2mm}
	\centerline{\includegraphics[width=0.49\textwidth]{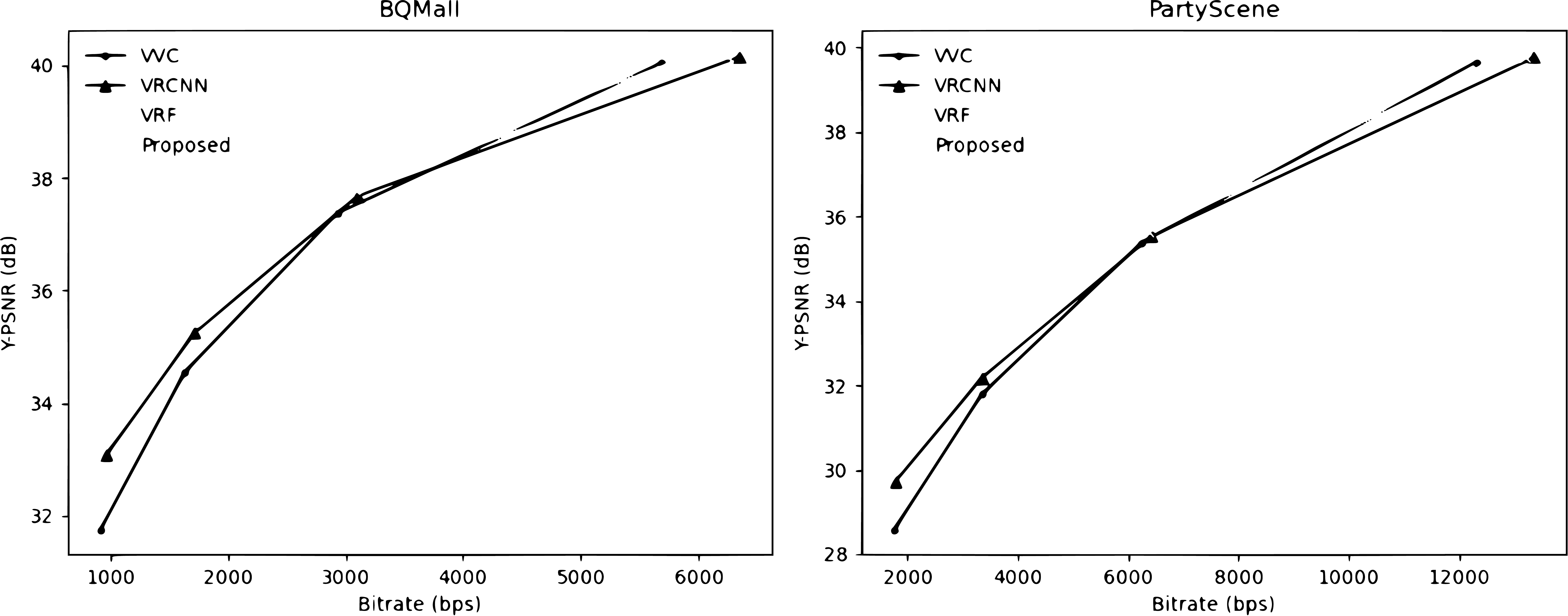}}
	\vspace{-3mm}
	\caption{Rate-distortion curves for two typical sequences.}
	\label{RDcurves}
	\vspace{-5mm}
\end{figure}

Fig.~\ref{example} shows the visual results of the sixth frame of the BQMall video sequence. We use VVC, VRCNN, VRF, and our proposed method to encode and the QP is 32. For verification purposes, some exemplar rate-distortion curves are shown in Fig.~\ref{RDcurves}. It can be observed that our scheme performs better at lower bit rates (high QP) than at higher bit rates (low QP). 

\begin{table}[t]
	\caption{BD-rate for Y component on common test sequences.}
	
	\begin{center}
		\scalebox{0.88}{
			\begin{tabular}{|c|c|c|c|c|}
				\hline
				\multirow{2}*{\textbf{Classes}} & \multirow{2}*{\textbf{Sequences}} & \multicolumn{3}{|c|}{\textbf{BD-Rate[\%]}}\\
				\cline{3-5}
				& & VRCNN\cite{VRCNN} & VRF\cite{EMC} & Proposed \\
				\hline
				
				\multirow{4}*{\tabincell{c}{ClassC \\ WVGA}} & BasketballDrill & -9.2 & -15.1 & -16.9 \\
				\cline{2-5}
				& BQMall & -4.2 & -9.1 & -13.1 \\
				\cline{2-5}
				& PartyScene & -2.1 & -6.3 & -8.9 \\
				\cline{2-5}
				& RaceHorses & -4.6 & -10.3 & -14.2 \\
				\hline
				
				\multirow{4}*{\tabincell{c}{ClassD \\ WQVGA}} & BasketballPass & -8.7 & -14.1 & -22.5 \\
				\cline{2-5}
				& BQSquare & 11.3 & 2.6 & -1.0 \\
				\cline{2-5}
				& BlowingBubbles & -11.4 & -16.1 & -20.3 \\
				\cline{2-5}
				& RaceHorses & -7.6 & -12.9 & -16.4 \\
				\hline
				
				\multirow{3}*{\tabincell{c}{ClassE \\ 720p}} & FourPeople & -4.2 & -5.2 & -6.1 \\
				\cline{2-5}
				& Johnny & -7.2 & -7.2 & -9.1 \\
				\cline{2-5}
				& KristenAndSara & -6.3 & -6.6 & -7.6 \\
				\hline
				
				\multirow{3}*{ClassF} & BasketballDrillText & -8.8 & -13.5 & -15.8 \\
				\cline{2-5}
				& ChinaSpeed & 9.8 & 7.6 & 7.2 \\
				\cline{2-5}
				& SlideEditing & 0.9 & 0.5 & 0.8 \\
				\hline
				
				\multirow{4}*{Average} & Class C & -5.0 & -10.2 & -13.3 \\
				\cline{2-5}
				& Class D & -4.1  & -10.1 & -15.1 \\
				\cline{2-5}
				& Class E & -5.9 & -6.3 & -7.6 \\
				\cline{2-5}
				& Class F & 0.6 & -1.8  & -2.6 \\
				\hline
				
				Overall & All & -3.6 & -7.1 & -9.7 \\
				\hline
				
		\end{tabular}}
		\label{tab2}
	\end{center}
	\vspace{-2mm}
\end{table}

\vspace{-2.5mm}
\section{Conclusion}
\vspace{-1mm}
This paper proposes a dilated convolution neural network, which is used to generate a deep reference picture for video coding. We use dilated convolution and inception module to increase the receptive field of the network so that it can obtain multi-scale feature map information, thereby improving the learning ability of the network. We also design a method to improve the coding efficiency of VVC. That is, the deep reference picture output by the network is employed to replace the original reference picture, and all the coding tests are conducted on the latest video coding platform VTM. Experimental results show that our scheme achieves significant bit saving compared to VVC. Future work will consider how to improve the transfer performance of the proposed model.

\bibliographystyle{IEEEbib}
\bibliography{IEEEexample}

\end{document}